\documentclass[prl,epsfig,twocolumn]{revtex4}
\usepackage{epsfig}
\usepackage{amstext}
\usepackage{amsmath}
\usepackage{algorithm}
\usepackage{algorithmic} 
\begin{document}
\title{Adaptive Cluster Expansion for Inferring Boltzmann Machines with Noisy Data}
\author{S. Cocco$^{1,2}$, R. Monasson$^{1,3}$}
\affiliation{
$^1$ The Simons Center for Systems Biology, Institute for Advanced Study, Einstein Drive, Princeton, NJ 08540, USA\\
$^2$ CNRS-Laboratoire de Physique Statistique de l'ENS, 24 rue Lhomond, 75005 Paris, France\\
$^3$ CNRS-Laboratoire de Physique Th\'eorique de l'ENS, 24 rue Lhomond, 75005 Paris, France}

\begin{abstract}
We introduce a procedure to infer the interactions among a set of binary variables, based on their sampled frequencies and pairwise correlations. The algorithm builds the clusters of variables contributing most to the entropy of the inferred Ising model, and rejects the small contributions due to the sampling noise. Our procedure successfully recovers benchmark Ising models even at criticality and in the low temperature phase, and is applied to neurobiological data. 
\end{abstract}
\maketitle


Understanding the correlated activity of complex, non-homogeneous multi-component systems is of fundamental importance in physics, biology, sociology, finance, ... A natural issue is to separate direct correlations (due to direct interactions) from network-mediated correlations. The Ising model, of ubiquitous importance in statistical physics, provides a natural framework to extract interactions from correlations \cite{maxent}, and was recently used for the analysis of neurobiological data \cite{bialek,marre,noi}. It is indeed the least constrained model capable of reproducing the individual and pairwise frequencies of a set of, say, $N$ binary-valued variables, $\sigma_i=0, 1$. In practice, these frequencies, $p_i$ and $p_{ij}$, are often estimated through empirical averages over a number of sampled configurations $\{\sigma_1^b,\sigma_2^b,\ldots,\sigma_N^b\}$, $b=1,\ldots,B$. The task then consists in inferring the parameters (fields $h_i$ and interactions $J_{ij}$) of the Ising model reproducing those data. From a mathematical point of view, one has to solve the $\frac 12N(N+1)$ implicit equations $p_i=\langle \sigma_i\rangle$ and $p_{ij}= \langle\sigma _i \sigma_j\rangle$ for the fields and interactions, where $\langle\cdot\rangle$ denotes the Gibbs average with Boltzmann factor $\exp \big(\sum _i h_i\sigma_i+\sum_{i<j} J_{ij}\sigma_i\sigma_j\big)$. 

Various approaches have been developed to solve the inverse Ising problem, called Boltzmann Machine (BM) in Machine Learning, including BM learning \cite{ackley}, mean field \cite{opper} and message-passing \cite{peli,mora} methods, and pseudo-likelihood algorithms \cite{wain}. Despite their specificities, those methods have in common to be efficient when the correlations, $c_{ij}=p_{ij}-p_ip_j$, are weak, and to perform badly when most pairs $(i,j)$ are strongly correlated, {\em e.g.} when the data are generated by a critical Ising model. Those examples seem to suggest that fast algorithms cannot infer BMs with long-range correlations \cite{montanari}.

However, the existence of a relationship between the presence of strong correlations in the 'direct' model and the intrinsic hardness of the inverse problem is questionable \cite{swendsen}. Let ${\bf p}=\{p_i,p_{ij}\}$, ${\bf J}=\{h_i,J_{ij}\}$, ${\bf \langle \boldsymbol\sigma \rangle}=\{\langle \sigma_i\rangle, \langle \sigma_i\sigma_j\rangle\}$ be the $\frac 12 N(N+1)$-dimensional vectors of, respectively, the measured frequencies, the interaction parameters and the Gibbs frequencies. We define the susceptibility and the inverse susceptibility matrices through, respectively, 
\begin{equation}\label{definvchi}
\boldsymbol\chi = \left. \frac{\partial {\bf \langle \boldsymbol\sigma \rangle}}{\partial {\bf J}}\right|_{\bf J} 
\quad \hbox{\rm and}\quad 
\boldsymbol\chi ^{-1} = \left. \frac{\partial {\bf J}}{\partial {\bf \langle \boldsymbol\sigma \rangle}} \right|_{\bf \langle \boldsymbol\sigma \rangle}\ .
\end{equation} 
$\boldsymbol\chi$ is attached to the direct model, and quantifies how the frequencies respond to a small change in the interaction parameters. $\boldsymbol\chi^{-1}$, which gives the response of the BM interaction parameters to a small change in the frequencies, is a natural characterization for the inverse problem. An essential point, which has received little attention in the context of BM so far, is that ${\boldsymbol \chi}^{-1}$ is generally much sparser and shorter-range than ${\boldsymbol \chi}$; evidence for this claim is reported below. Even if strong responses (and correlations) pervade the system, each BM interaction parameter may mostly depend on a small (compared to $N$) number of frequencies $\bf p$. Interestingly, the short-rangedness of ${\boldsymbol \chi}^{-1}$ makes the inference not only possible but also meaningful, as experiments generally probe limited parts of larger systems.

In this letter, we present a method for inferring BM, exploiting this notion of limited dependence. The interaction network is progressively unveiled, through a recursive processing of larger and larger subsets of variables, which we call clusters. To each cluster $\Gamma$ is associated an entropy $\Delta S(\Gamma)$, which assesses how much the cluster is relevant to infer the BM. Clusters such that $|\Delta S(\Gamma)|<\Theta$, where $\Theta$ is a fixed threshold are discarded; the other clusters are kept and recursively used to generate larger clusters. Threshold $\Theta$ must be large enough to avoid overfitting of the data corrupted by the sampling noise (finite $B$) and small enough in order not to miss important components of the interaction network. Contrary to conventional cluster expansions \cite{noi,peli}, the number, size, and composition of the clusters automatically adapt to the data, and, rather than the sole value of $N$, determine the running time of the algorithm. Pseudo-codes intended for the practical implementation of our algorithm are given in Supplemental Material \cite{si}.

Our starting point is the Legendre transform of the partition function $Z({\bf J})$ (sum of the Boltzmann factors) of the Ising model, 
\begin{equation}
\label{entroising} 
S ({\bf p})= \min_{\bf J} \big[\log Z({\bf J}) - {\bf p}\cdot {\bf J} \big] \ ,
\end{equation}
where $\cdot$ denotes the dot product; it is the cross entropy between the sampled distribution and the best BM or, equivalently, the negative of the maximum log-likelihood of the parameters $\bf J$ given the data $\bf p$ \cite{cover}. Let us define $S_0({\bf p}) = \frac 12\log \hbox{\rm det} (\hat c_{ij})$, where $\hat c_{ij}=c_{ij}/[p_i(1-p_i)p_j(1-p_j)]^{\frac 12}$. We now formally write, for given $\bf p$, 
\begin{equation}\label{recur-entro} 
S-S_0= \sum _i \Delta S{(i)} +
 \sum _{i<j} \Delta S{(i,j)} + \sum _{i<j<k} \Delta S{(i,j,k)} +\ldots\ ,
\end{equation} 
where the sums run over every subset (cluster) of the $N$ variables. The choice of expanding $S-S_0$ rather than $S$ will be explained later. According to (\ref{recur-entro}) for $N=1$, $\Delta S{(i)}$ is the entropy of a single spin with average value $p_i$.  Using (\ref{recur-entro}) again for $N=2$, we find that $\Delta S{(i,j)}$ equals the loss in entropy when imposing the constraint $\langle\sigma_i\sigma_j\rangle =p_{ij}$ to a system of 2 spins with fixed magnetizations, $\langle \sigma_i\rangle=p_i$, $\langle \sigma_j\rangle=p_j$, minus the contribution $\frac 12 \log (1-\hat c^2_{ij})$ coming from $S_0$. A recursive use of (\ref{entroising}) and (\ref{recur-entro}) for increasing $N$ allows us to calculate $\Delta S({\Gamma})$ for larger and larger clusters $\Gamma=(i_{1},i_{2},\ldots ,i_{K})$. The maximal cluster size, say, $K=20$, is set by the computational hardness of obtaining $S$ from (\ref{entroising}). Note that $\Delta S(\Gamma)$ is a function of the individual and pairwise frequencies of the spins in $\Gamma$ only.

\begin{figure}[t]
\begin{center}
\epsfig{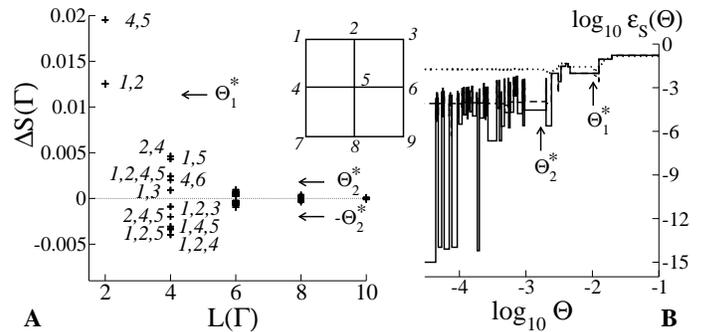}
\caption{Exact cluster enumeration for a $3\times 3$ grid, with coupling $J=1.778$ (units of $k_BT$), corresponding to the critical value for an infinite grid \cite{gridtc}. {\bf A.} Cluster-entropies $\Delta S({\Gamma})$ vs. length $L(\Gamma)$ of the interaction path for perfect sampling; representative $\Gamma$ are shown for $L=2,4$ (labels refer to the grid). {\bf B.} error $\epsilon_S$  vs. $\Theta$ for perfect sampling ($B=\infty$, full curve), and two random samples with $B=10^7$ (dashed) and $B=4500$ (dotted curve) configurations. The accuracy on $\epsilon_S$ and each $\Delta S$ is $\sim 10^{-15}$.}
\label{fig-grid}
\end{center}
\end{figure}

To illustrate the properties of the cluster expansion (\ref{recur-entro}) consider the 2D-Ising model on a $M\times M$ grid (Fig.~\ref{fig-grid}), in the absence of sampling noise ($B=\infty$). Enumerations of the $2^{M^2}$ spin configurations allow us to calculate the frequencies ${\bf p}=\langle \boldsymbol\sigma\rangle$ and the cluster-entropies $\Delta S(\Gamma)$ exactly for small values of $M$ (Fig.~\ref{fig-grid}A). The entropy of the clusters $\Gamma$ decreases exponentially with the length $L(\Gamma)$ of the shortest closed interaction path joining the spins in $\Gamma$, {\em e.g.} $L(1,2)=2$, $L(1,3,6)=6$. The entropies of clusters sharing a common interaction path (and the same $L$) have alternating signs, depending on the parity of the cluster size (Fig.~\ref{fig-grid}A); their sum is much smaller (in absolute value) than any cluster-entropy taken separately \footnote{This (partial) cancellation property ensures that $S$ is extensive in $N$.}. Figure~\ref{fig-grid}B shows the error $\epsilon _S$ on the entropy, when all cluster-entropies smaller than $\Theta$ are discarded. $\epsilon_S$ exhibits lower and lower plateaus, separated by higher barriers as the threshold $\Theta$ is decreased. The first low plateau, $\epsilon_S \simeq.002$, takes place at $\Theta^{*}_1= .012$, when all nearest-neighbor clusters ($L=2$) are selected. The second and lower plateau, $\epsilon_{S} \simeq 5\,10^{-6}$, is reached for $\Theta^{*}_2=0.002$, after all clusters  with $L=4$ are taken into account. Barriers in between plateaus correspond to values of $\Theta$, for which the truncation interrupts the summation (and partial cancellation) of all the clusters sharing an interaction path; the error on the entropy is then $\epsilon_S\sim \Theta$.

\begin{figure}[t]
\begin{center}
\epsfig{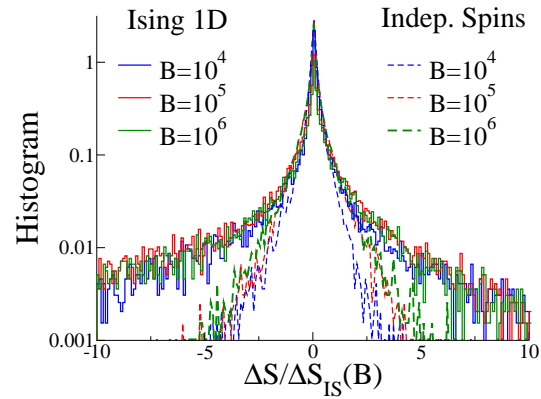}
\caption{Histograms of $\Delta S(i_{1},i_{2},i_{3})$ for the 1D-Ising ($J=4,h=-5$, units of $k_BT$) and Independent Spin (IS) models, for $N=50$ and three values of $B$. The distributions collapse onto each other after rescaling by the standard deviation of the IS cluster-entropies, $\Delta S_{IS}(B)\simeq \frac{3(2p-1)^2} {2p(1-p)} (\frac 3B)^{\frac 52} +O(\frac 1{B^3})$. Universality at small $\Delta S$ holds for larger cluster sizes ($=3$ here). Large-$\Delta S$ tails are not universal (not shown), and are specific to the interaction network, see Fig.~\ref{fig-grid}A.}
\label{fig-histo}
\end{center}
\end{figure}

Let us turn to the case of imperfect sampling (finite $B$). The measured correlations, $c_{ij}=p_{ij}-p_ip_j$, differ from the Gibbs correlations, $\langle \sigma_i\sigma_j\rangle - \langle \sigma_{i}\rangle\langle \sigma_{j}\rangle $, by random fluctuations of amplitude $\nu=O(B^{-\frac12})$. Those fluctuations do not affect much the largest correlations and the largest cluster-entropies. However, for the pairs $i,j$ with weak Gibbs correlations ($<\nu$ in absolute value), the measured correlations are dominated by the noise. This fact has two consequences. First, the norm of the 2-point susceptibility, $\displaystyle{|\chi _{2}| =\frac 1N \sum_{i,j} c_{ij}^2 \sim N \nu^2}$ is extensive: overfitting makes the inferred Ising model look like critical. Secondly, the distribution of the cluster entropies is universal for $\Delta S \to 0$ and $N\to\infty$: it coincides with the distribution for a system of Independent Spins, with the same $p_i$'s as the original system, and the same number $B$ of sampled configurations (Fig.~\ref{fig-histo}). The presence of this universal, noisy peak justifies the introduction of a threshold $\Theta$ and sets a lower bound to its value. Figure~\ref{fig-grid}B shows that the error $\epsilon _S$ behaves as in the perfect sampling case for large $\Theta$ and saturates at low $\Theta$ as expected. Again, the entropy is accurately estimated by taking into account only the top cluster-entropies, associated to the dominant interaction paths on the lattice.


Systematic enumeration of clusters is not possible for large systems. The example above suggests a fast, recursive procedure to build up clusters of increasing sizes, whose principle is based on the existence of paths of strong interactions connecting the spins. First we calculate the entropies associated to the $N$ clusters with $K=1$ spin. Then, two clusters $\Gamma _1$ and $\Gamma _2$ of size $K$ can be merged to give birth to the cluster $\Gamma = \Gamma_1 \cup \Gamma_2$ of size $K+1$ if $\Gamma_1$ and $\Gamma _2$ have exactly $K-1$ common spins, and if $|\Delta S(\Gamma)| >\Theta$. The underlying principle is, again, that the building-up prescription should be compatible with the existence of a path of strong interactions connecting the spins, and that clusters with low entropies can be discarded. Each time a new cluster $\Gamma$ is created and selected we store its contributions to the entropy, $\Delta S(\Gamma)$ and to the interaction parameters, $\Delta {\bf J}({\Gamma}) = - \frac{d}{d {\bf p}} \Delta S(\Gamma)$. The procedure naturally stops when no cluster of larger size can be built through the recursion. The sums of $\Delta S(\Gamma)$ and $\Delta {\bf J}({\Gamma})$ over the selected clusters, added to, respectively, $S_{0}$ and ${\bf J}_{0}=- \frac{d}{d {\bf p}} S_0$, are our approximations for the entropy and the interactions of the BM.

\begin{figure}[t]
\begin{center}
\epsfig{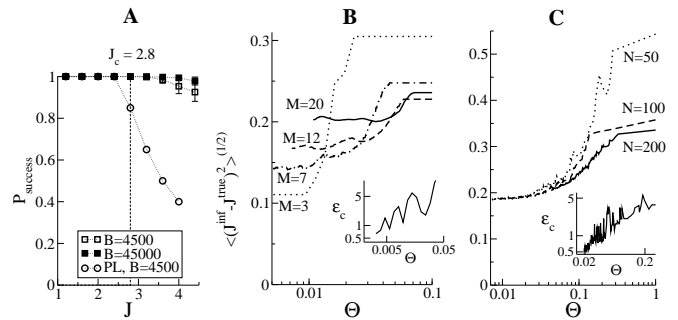}
\caption {{\bf A.} Fraction of non-zero interactions recognized by our procedure (squares) and by the PL algorithm (circles, from \cite{montanari}) vs. intensity $J$ of couplings on a $7\times 7$ grid with 30\% dilution; similar performances are obtained for $20\times 20$ grids. The critical coupling is $J_c\simeq 2.8$ (units of $k_BT$) \cite{gridtc}.  {\bf B.} Error on the inferred couplings vs. $\Theta$ for $M\times M$ grids at 'criticality' ($J_c=1.778$, no dilution) and $B=4500$; the dependence on $M$ is reduced with periodic boundary conditions (not shown).  Inset: $\epsilon_c$ vs. $\Theta$ for $M=7$. {\bf C.} same as {\bf B} for the Viana-Bray spin glass model \cite{vb} (connectivity 5 and random $J_{ij}$, uniform in $[-J^0;J^0]$; $J^0=4$ is larger than the spin glass critical coupling, $J^0_c \simeq 3.5$ \cite{vb}). Inset: $\epsilon_c$ vs. $\Theta$ for $N=50$.}
\label{fig-syn}
\end{center}
\end{figure}

We now report the tests of the above inference algorithm on synthetic data generated from Ising models with known couplings. First we consider dilute ferromagnets on 2D-grids of sizes $M\times M$; BM learning is hindered by the huge thermalization time at low temperature, mean-field and message-passing methods are not expected to be efficient on such loopy lattices and the Pseudo-Likelihood (PL) algorithm of \cite{wain} fails outside the paramagnetic phase, even for $M=7$ \cite{montanari}. Our algorithm successfully retrieves the network of interactions at the critical point, in the low temperature phase, and for much larger sizes (Fig.~\ref{fig-syn}A). As $\Theta$ is lowered the error on $J_{ij}$ first decreases and then saturates to a value close to the Cram\'er-Rao bound, $\sqrt{\frac 1{B}\, {\boldsymbol\chi}^{-1}_{ij,ij}}$ \cite{cover} (Fig.~\ref{fig-syn}B). At the cross-over threshold the largest selected clusters have size 4, while $\xi\sim M$ as the system is critical (Fig.~\ref{fig-syn}B). The running time of the algorithm (at the cross-over $\Theta$) is $\sim 10$ millisec on one core of an AMD Opteron dual-core processor at 3 Ghz. The inference algorithm is also applied to glassy frustrated Ising models \cite{vb}, of various sizes $N$ (Fig.~\ref{fig-syn}C). Performances do not seem to worsen as $N$ increases. 

To better understand the saturation of the error and the quality of the inference we compare the difference $\delta {\bf p}$ between the frequencies calculated from the inferred BM, ${\bf \langle \boldsymbol\sigma \rangle}$ \footnote{These can be calculated using Monte Carlo simulations, or a cluster expansion (this time, for the direct problem) with a threshold; details will be given elsewhere.}, and the measures, ${\bf p}$, to the fluctuations expected from the sampling of $B$ configurations at equilibrium. The variance of these fluctuations are the diagonal elements of $\boldsymbol\chi$, divided by $B$. An estimate of the relative error for the one-site frequencies is thus $\epsilon _p = \sqrt{\frac BN \sum _{i}\frac { (\delta p_{i})^2}{\chi _{i,i}}}$; a similar expression can be written for the error on the correlations, $\epsilon _c$. Values of $\epsilon \gg 1$ signal a poor inference, while overfitting corresponds to $\epsilon \ll 1$. This criterion is justified if the Gibbs fluctuations are comparable to the error bars that can be computed using statistical methods such as bootstrap. We find that $\epsilon_p$ and $\epsilon_c$ are close to 1 at the cross-over threshold for which the error on the couplings saturates (Insets of Fig.~\ref{fig-syn}B\&C). Lowering $\Theta$ further reduces $\epsilon_p,\epsilon_c$, but does not increase the accuracy on the interactions and is merely an overfitting of the data.

\begin{figure}[t]
\begin{center}
\epsfig{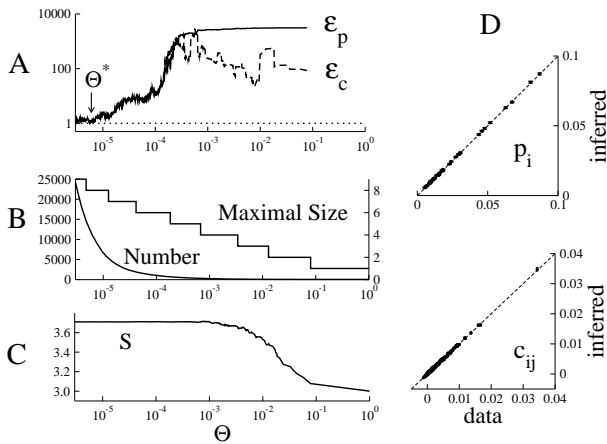}
\caption{Performance of the inference procedure as a function of the threshold $\Theta$. 
{\bf A.} errors $\epsilon _p$ and $\epsilon_c$; 
{\bf B.} number (bottom, left scale) and maximal size (top, right scale) of clusters;
{\bf C.} entropy $S(\Theta)$. {\bf D.} reconstructed vs. experimental $p_i$ and $c_{ij}$ for $\Theta^*$ (error bars are calculated from $\chi$).  Spin values are $\sigma_i=1$ if cell $i$ is active in a 20 msec-time bin, 0 otherwise \cite{bialek}. Note that, while the correlations $c_{ij}$ are small, the ratios $c_{ij}/(p_ip_j)$ are of the order of unity, and so are the inferred couplings $J_{ij}$.}
\label{fig-errorvsT}
\end{center}
\end{figure}

The running time of our algorithm depends on the complexity of the underlying interaction network rather than on the system size. We analyze in Fig.~\ref{fig-errorvsT} a 3180 second-long recording of the retinal activity of a salamander, previously studied in \cite{bialek} using BM learning ($N_1=40$ cells). As $\Theta$ is lowered, the number of selected clusters and their maximal size increase, and the entropy $S$ reaches a plateau (Fig.~\ref{fig-errorvsT}B\&C). For $\Theta^*\simeq 6\, 10^{-6}$, the errors are $\epsilon \sim 1$, and the inferred Ising model reproduces the frequencies and correlations (Fig.~\ref{fig-errorvsT}D). We have also applied our algorithm to recordings of other neurobiological systems, including the cortical activity of $N_2=37$ cells in a behaving rat \cite{Pey09} (not shown). While the amplitudes of the interactions found with both data sets are similar, the maximal size of the selected clusters, which is a measure of the neighborhood of a cell on the interaction network, is much smaller for the cortical recording ($=3$) than for the retinal activity ($=8$). The lower complexity of the inferred network results in a lower CPU time ($t_2\simeq .1$ sec vs. $t_1\simeq 5$ min on the computer above), in spite of $N_2 \simeq N_1$.


While expanding $S$ alone in (\ref{recur-entro}) would be possible, the cluster-entropies $|\Delta S_{\Gamma}|$ produced by the expansion of $S-S_{0}$ are generally smaller \cite{diag}. Therefore, less clusters are needed to achieve an accurate inference, and the fluctuations of $\epsilon_S$ (Fig.~\ref{fig-grid}B) and of $\epsilon_p,\epsilon_c$ (Figs.~\ref{fig-syn}B\&C and \ref{fig-errorvsT}A) are smaller, see discussion about barriers above.  As $S_{0}$ coincides with $S$ for mean field models when $N\to\infty$ \cite{opper}, it is a good starting point for the expansion even for systems with rather dense  and weak interaction networks. In the case of severe undersampling, regularized versions of $S_0$ including a penalty over the couplings based on the $L_2$ (\cite{si}) or the $L_1$ \cite{lamfan} norm can be used. Note that $({\bf J}_0)_{ij} \propto - (\hat {\bf c}^{-1})_{ij}$ is regular even at criticality, {\em i.e.} even if $\hat{\bf c}$ has a diverging eigenvalue. 

Our work suggests that the BM problem can be solved efficiently even when data exhibit strong correlations. The contribution to $\boldsymbol\chi^{-1}$ due to a cluster $\Gamma$, $-\frac{\partial^2 \Delta S_\Gamma}{\partial {\bf p}\partial{\bf p}}$, is highly sparse since ${\Delta S}_\Gamma$ depends on a few frequencies only. The success of our algorithm relies on the property that $\boldsymbol\chi^{-1}$ can be accurately approximated by such an expansion (while $\boldsymbol\chi$ cannot). We now list four examples for which this property holds. In the 1D-Ising model, ${\boldsymbol\chi}_{ij}^{-1}$ is of finite-range when $J_{ij}$ couples nearest neighbours only, and decays exponentially with $|i-j|$ in the presence of longer-range interactions \cite{percus}. Next, consider the $O(m)$ model, where the binary spins $\sigma_i$ are replaced with $m$-dimensional spins $\boldsymbol \sigma _i$ of fixed norms, with interactions $J_{ij}$ and zero fields. The model is exactly solvable in the $m\to\infty$ limit, with the result $(\boldsymbol \chi^{-1})_{ij,kl}= J_{ik} J_{jl } +J_{il} J_{jk}$ (diagonal elements $J_{ii}$ enforce the constraints on $|\boldsymbol \sigma _i|$). If ${\bf J}$ is sparse, so is ${\boldsymbol\chi}^{-1}$, even if all correlations are strong. In liquid theory, the Ornstein-Zernike direct correlation function, a quantity closely related to ${\boldsymbol \chi}^{-1}$, is widely believed to be short-range \cite{hansen}; this property is used in closure schemes, {\em e.g.} Percus-Yevick, to obtain the equation of state. Even at the critical point of a ferromagnet \cite{swendsen} the response of the field $h_i$ to changes in the magnetizations $m_j$ of spins at distance larger than $R$, $\int _{r>R} dr\, |{\boldsymbol \chi}^{-1}(r)|\sim R^{-(3-\eta)}$, quickly decays with $R$ \cite{hansen}. Intuitively, the $O(N^2)$ correlations contain a highly redundant information about the $O(N)$ non-zero couplings which have generated them. This redundancy is at the origin of the 'locality' of ${\boldsymbol \chi}^{-1}$ and of the cancellation property of the cluster-entropies. \\
We thank D. Chatenay, D. Huse, J.~Lebowitz, S. Leibler, A. Montanari and V. Sessak for very useful discussions.


\end{document}